\pdfoutput=1

\documentclass[11pt,a4paper]{article}
\usepackage{jcappub}

\usepackage{bm}
\usepackage{epsfig}
\usepackage{citesort}
\usepackage{graphicx}
\usepackage{amssymb}
\usepackage{color}
\usepackage{subfigure}

\newcommand{\be}{\begin{equation}}
\newcommand{\ee}{\end{equation}}
\newcommand{\bea}{\begin{eqnarray}}
\newcommand{\eea}{\end{eqnarray}}

\newcommand{\beq}{\begin{eqnarray}}
\newcommand{\eeq}{\end{eqnarray}}

\def\nue{\mathrel{{\nu_e}}}

\def\barnue{\mathrel{{\bar \nu}_e}}

\def \lta {\mathrel{\vcenter{\hbox{$<$}\nointerlineskip\hbox{$\sim$}}}}
\def \gta {\mathrel{\vcenter{\hbox{$>$}\nointerlineskip\hbox{$\sim$}}}}

\def\t13{\mathrel{{\theta_{13}}}}
\def\y12{\mathrel{{\tan^2 \theta_{12}}}}
\def\c2{\mathrel{{\chi^2 }}}

\def\msun{\mathrel{{M_\odot}}}

\newcommand{\n}{neutrino}
\newcommand{\ns}{neutrinos}

\newcommand{\sn}{SN}
\newcommand{\sne}{supernovae}

\newcommand{\df}{DSNB}
\newcommand{\snr}{SNR}
\newcommand{\sfr}{SFR}

\newcommand{\sk}{Super-Kamiokande}

\begin{document}

%%%%%%%%%%%%%%%%%%%%%%%%%%%%%%%%%%%%%%%%%%%%%%%%%%%%%%%%%%%%%%%%%%%%%%
% Frontpage %%%%%%%%%%%%%%%%%%%%%%%%%%%%%%%%%%%%%%%%%%%%%%%%%%%%%%%%%%
%%%%%%%%%%%%%%%%%%%%%%%%%%%%%%%%%%%%%%%%%%%%%%%%%%%%%%%%%%%%%%%%%%%%%%

\subheader{\hfill MPP-2012-90}

\title{Diffuse supernova neutrinos: oscillation effects, stellar cooling and progenitor mass dependence}

\author[a,b]{Cecilia Lunardini}
\author[c]{and Irene Tamborra}

\affiliation[a]{Arizona State University, Tempe, AZ, 85287-1504, USA}
\affiliation[b]{RIKEN BNL Research Center, Brookhaven National Laboratory, Upton, NY 11973, USA}

\affiliation[c]{Max-Planck-Institut f\"ur Physik
(Werner-Heisenberg-Institut)\\
F\"ohringer Ring 6, 80805 M\"unchen, Germany}

\emailAdd{Cecilia.Lunardini@asu.edu}
\emailAdd{tamborra@mpp.mpg.de}

\abstract{ 
We estimate the diffuse supernova neutrino background (DSNB) using the recent progenitor-dependent, long-term supernova 
simulations from the Basel group and including \n\ oscillations at several post-bounce times. 
Assuming multi-angle matter suppression of collective effects during the accretion phase, we find that oscillation 
effects are dominated by the matter-driven MSW resonances, while \n-\n\ collective effects contribute at the $5$--$10 \%$ level.  
The impact of the \n\ mass hierarchy, of the time-dependent \n\ spectra and of the diverse progenitor star population  
is  $10 \%$ or less, small compared to the uncertainty of at least $25 \%$ of the normalization of the supernova rate. 
Therefore, assuming that the sign of the neutrino mass hierarchy will be determined within the next decade, the future detection of the DSNB will deliver approximate information on the MSW-oscillated \n\ spectra. 
With a reliable model for \n\ emission, its detection will be a powerful instrument to provide complementary information on the star formation rate
and for learning about stellar physics.} 
\maketitle

%%%%%%%%%%%%%%%%%%%%%%%%%%%%%%%%%%%%%%%%%%%%%%%%%%%%%%%%%%%%%%%%%%%%%%
\section{Introduction}                        \label{sec:introduction}
%%%%%%%%%%%%%%%%%%%%%%%%%%%%%%%%%%%%%%%%%%%%%%%%%%%%%%%%%%%%%%%%%%%%%%

The Diffuse Supernova Neutrino Background (DSNB) is the flux of neutrinos and antineutrinos emitted by all core-collapse supernovae (SN) in our universe.  Isotropic and stationary, it is a guaranteed signal that will give us a unique image of the \sn\ population of the universe and of its history all the way to redshift $\sim 1$ or so. This image can be a precious alternative, or a unique complement, to a high statistics detection of an individual nearby \sn\ which is a rare event ($\sim 1$-$3$ SNe per century in our galaxy~\cite{Arnaud:2003zr,Ando:2005ka,Horiuchi:2011zz,Botticella:2011nd}). 
Although the DSNB has not been detected yet, 
its discovery prospects are excellent. Gadolinium enriched  Super-Kamiokande is expected  to detect the DSNB at a level of few events per year~\cite{Beacom:2003nk,Horiuchi:2008jz} and  tens to hundreds of events are predicted  for the next generation of larger scale, $0.1$--$1$~Mt mass detectors, running for  a few years~\cite{deBellefon:2006vq,Nakamura:2003hk}.

The DSNB depends on the cosmological SN rate, the neutrino fluxes  and the oscillation physics. 
Its prediction is a challenging problem because of the still sparse astrophysical observations of \sne, the incomplete picture of \n\ oscillation physics, and several obstacles --- of theoretical and computational nature --- in the way of fully modeling the \n\ emission from core collapse.  Due to these difficulties, a number of simplifying assumptions were adopted to model the \df, typically the same neutrino spectra were used for different post-bounce times and  for all progenitors and only approximate neutrino oscillation physics was considered (see~\cite{Lunardini:2010ab,Beacom:2010kk} for reviews on the topic). 
Only recently efforts were developed to include more accurate astrophysical information and 
 oscillations driven by \n-\n\ forward scattering~\cite{Ando:2004hc,Galais:2009wi,Chakraborty:2008zp,Chakraboty:2010sz}.

From the perspective of neutrino oscillation physics, lately new advances have taken place on the experimental and theoretical front. The elusive mixing angle $\theta_{13}$ has been measured~\cite{An:2012eh,Ahn:2012nd}, leaving us only with open questions on the \n\ mass hierarchy  and the CP-violating phase (although the latter does not affect neutrino oscillations in supernovae~\cite{Gava:2008rp}).  Concerning neutrino oscillations in supernovae from the theoretical perspective,  the matter enhanced flavor conversion picture~\cite{wolf,Wolfenstein:1977ue} has been  completed  by including  
neutrino-neutrino interactions, responsible for the non-linear neutrino flavor evolution (see~\cite{Duan:2010bg} for a review on the topic). 
The present sketch is that, during the accretion phase, 
 total (or partial) multi-angle matter suppression of collective oscillations is 
expected~\cite{Dasgupta:2011jf,Mirizzi:2011tu,Saviano:2012yh,Chakraborty:2011gd,Chakraborty:2011nf,Sarikas:2011am}; while, during the cooling phase, multiple spectral splits 
occur~\cite{Dasgupta:2009mg,Fogli:2009rd} and  Mikheev-Smirnov-Wolfenstein (MSW) resonances are expected at larger radii~\cite{wolf,Wolfenstein:1977ue}. 
Moreover, most recent numerical simulations of core-collapse SN provide information about \n\ spectra up to $\sim 10$ s post-bounce~\cite{Fischer:2009af,Mueller:2012is} and for different progenitor masses~\cite{Fischer:2009af}.  Therefore the time is mature to include these updates in the estimate of the \df\  and to assess the impact of each in comparison with the many uncertainties, experimental and theoretical, that affect the diffuse flux itself.  This is the goal of this paper. 
We model the \df, as consistently as possible, in the framework of the Basel model~\cite{Fischer:2009af}. We consider oscillation physics for different post-bounce times for both neutrino mass hierarchies, including the multi-angle treatment of $\nu$--$\nu$ interactions.

This work is organized as follows. In Section~\ref{sec:generalities} we 
discuss the cosmological supernova rate, the adopted supernova models for different progenitor masses 
and we give a brief overview on neutrino oscillation physics in supernovae. In Section~\ref{sec:models} we
discuss the impact of neutrino oscillations on the neutrino time-integrated flux 
for a fixed supernova mass. In Section~\ref{sec:dsnb}, we present our results on the DSNB.  
Conclusions and perspectives are illustrated in Section~\ref{sec:conclusions}.

%%%%%%%%%%%%%%%%%%%%%%%%%%%%%%%%%%%%%%%%%%%%%%%%%%%%%%%%%%%%%%%%%%%%%%
\section{Input physics}                        
\label{sec:generalities}
%%%%%%%%%%%%%%%%%%%%%%%%%%%%%%%%%%%%%%%%%%%%%%%%%%%%%%%%%%%%%%%%%%%%%%
In this Section, the cosmological SN rate and the reference neutrino signal for different SN masses are discussed. 
We also introduce the neutrino mass-mixing parameters and the quantum kinetic equations describing the neutrino
oscillations in supernovae. 

%%%%%%%%%%%%%%%%%%%%%%%%%%%%%%%%%%%%%%%%%%%%%%%%%%%%%%%%%%%%%%%%%%%%%%
\subsection{Diffuse \sn\ \n\ flux}                        
\label{sec:dsnbgeneral}
%%%%%%%%%%%%%%%%%%%%%%%%%%%%%%%%%%%%%%%%%%%%%%%%%%%%%%%%%%%%%%%%%%%%%%

Simply put, the \df\ offers us an image of the entire SN  population of the universe.  It reflects the demographics of this population: how it is  distributed in space, its history, the diverse sub-types that contribute to it with their different \n\ emissions.  Besides, the image we get is influenced by the neutrinos' own history, as they travel cosmic distances before they reach us. Therefore the diffuse flux of a given \n\ flavor  depends on many physical quantities of different origin.
 
 \begin{itemize}
 \item The flux produced by an individual SN, which depends on the dynamics of the core collapse and is different for different masses of the progenitor star. 
\item Propagation effects: namely flavor oscillations, which depend on the \n\ masses and mixings, 
 and the redshift of energy; for a \n\ produced at redshift $z$ with energy $E^\prime$, the  observed energy is $E = E^\prime/(1+z)$.     
\item  The comoving SN \ rate (SNR), $\dot{\rho}_{\rm SN}(z,M) $,  per unit of redshift and per unit of progenitor mass $M$. It is defined in the mass interval between $M _0 \simeq 8 \msun$ (the minimum necessary to have a core-collapse supernova) and a maximum $M_{\rm max}\sim 125 \msun$, the tentative upper limit for the occurrence of normal core-collapse supernovae (as opposed to pair instability ones or black-hole forming events)~\cite{Ando:2004hc}.
In terms of redshift, the \snr\ is largest between $z=0$ and $z_{\rm max} \simeq 5$, reflecting the period of most intense star formation activity (see Sec.~\ref{sec:CCSFR}). Note that  we neglect the contribution  due to failed supernovae~\cite{Lunardini:2009ya}.  
\end{itemize}

With these ingredients, the diffuse  flux for each flavor $\nu_\beta$ ($\beta = e, \bar{e}, \mu$ or~$\tau$) can be written as~\cite{Ando:2004hc}:
\begin{equation}
\Phi_{\nu_\beta}(E)=\frac{c}{H_0} \int_{M_0}^{M_{\rm max} } dM \int_{0}^{z_{\rm max}}  dz\ \frac{\dot{\rho}_{SN}(z,M) F_{\nu_\beta}(E',M)}{\sqrt{\Omega_M (1+z)^3+\Omega_\Lambda}}~,
\label{phiDSNB}
\end{equation}
where $c$ is the speed of light and $H_0 = 70\ \rm{km}~\rm{s}^{-1}~\rm{Mpc}^{-1}$ the Hubble constant;   $\Omega_M = 0.3$  and $\Omega_\Lambda = 0.7$ are  the fractions of the cosmic energy density in matter and dark energy respectively and $F_{\nu_\beta}(E^\prime,M)$ is the oscillated $\nu_\beta$ flux 
for a \sn\ with progenitor mass $M$ 
(see Sec.~\ref{sec:models}).  As we will see, the \df\  is dominated by the contribution of the closest ($z \lta 1$) and least massive ($M \sim M_0$) stars, and depends only weakly on $M_{\rm max}$ and $z_{\rm max}$. 

Although the \df\ has not been observed yet, interesting upper limits exist. The most stringent is on the $\barnue$ component of the flux, from a search of inverse beta-decay events at \sk, above 17.3 MeV threshold: $\phi_{\barnue} \lta 2.8$--$3.0~{\rm cm^{-2} s^{-1}} $ at 90\% C.L.~\cite{Bays:2011si,Ikeda:2007sa}. This bound is generally consistent with predictions, excluding scenarios where multiple parameters conspire to generate a particularly large flux~\cite{Yuksel:2007mn}. 
Because the search at Super-Kamiokande is background-dominated,  any substantial improvement on it will require  better background subtraction.   Methods involving water with Gadolinium addition~\cite{Beacom:2003nk}, liquid Argon~\cite{Cline:2006st,Ereditato:2005yx,Rubbia:2009md}, and liquid scintillator~\cite{Wurm:2011zn} are especially promising.  Of these, detectors of Megaton class will have the further advantage of high statistics, yielding up to hundreds of events a year from the \df\ (see e.g.,~\cite{deBellefon:2006vq,Nakamura:2003hk}).

%%%%%%%%%%%%%%%%%%%%%%%%%%%%%%%%%%%%%%%%%%%%%%%%%%%%%%%%%%%%%%%%%%%%%%
\subsection{Cosmological supernova rate}                        
\label{sec:CCSFR}
%%%%%%%%%%%%%%%%%%%%%%%%%%%%%%%%%%%%%%%%%%%%%%%%%%%%%%%%%%%%%%%%%%%%%%

Considering that \sn\ progenitors are very short lived, the \snr\ is proportional to the Star Formation Rate (\sfr), $\dot{\rho}_{\star}$, defined as the mass that forms stars per unit time per unit volume. The relationship between the \snr\ and \sfr\ is given by the  Initial Mass Function (IMF), $\eta(M)\propto M^{-2.35}$~\cite{Salpeter:1955it}, which describes the mass distribution of stars at birth: 
\be
\dot{\rho}_{SN}(z,M) = \frac{\eta(M)}{\int_{0.5 \msun}^{M_{\rm max} }dM\ M \eta(M)}   \dot{\rho}_{\star}(z)\ .
\label{snr}
\ee
Recent  analyses of \snr\ and \sfr\ measurements~\cite{Hopkins:2006bw} show that a piecewise parametrization of $\dot{\rho}_{\star}$  is adequate~\cite{Hopkins:2006pr}: 
\begin{eqnarray}
\dot{\rho}_{\star} \propto\Bigg\{ \begin{array}{lc}{(1+z)^{\delta}}\ \ \ \ \ \ \ z<1 \\
{(1+z)^{\alpha}}\  \ \ \ 1<z<4.5\ .\\
{(1+z)^{\gamma}}\ \  4.5<z
\end{array} 
\label{snrparam}
\end{eqnarray}
Here we adopt this function, with $\delta,\alpha,\gamma$ and the normalization fixed at the best fit values~\cite{Hopkins:2006pr}: $\delta=3.28$, $\alpha=-0.26$, $\gamma=-7.8$ and $\int_{M_0}^{M_{\rm max} }dM\  \dot{\rho}_{SN}(0,M)  = 1.5 \times 10^{-4}~{\rm Mpc^{-3} yr^{-1}}$ (integrated \snr\ at the present epoch). 
Note that, due to the redshift of energy, \ns\ of higher redshift accumulate at lower energies, so that in the energy window relevant for experiments ($11~{\rm MeV } \lta E \lta 40~{\rm MeV } $~\cite{Horiuchi:2008jz,Bays:2011si}) the diffuse flux is dominated by the low $z$ contribution, $z\lta 1$. Therefore, its dependence on $\alpha$ and $\gamma$ is weak.  The flux is also dominated by the lower mass stars, considering the fast decline of the IMF with $M$, so that there is a strong dependence on $M_{0}$, but a weak one  on the high cutoff $M_{\rm max}$. 
 
Let us now comment on the existing measurements of  the \snr\ and their uncertainties, which are dominated by normalization errors~\cite{Hopkins:2006bw,Horiuchi:2011zz}. 
Perhaps the most precise way to measure the \snr\ is from data on the \sfr, via Eq.~(\ref{snr}). The cosmic star formation history as a function of the redshift  is known from data in the ultraviolet and far-infrared~\cite{Hopkins:2006pr}.  Constraints are stronger at $z\lta 1$, while observations at higher redshift have larger errors.  Following Ref.~\cite{Hopkins:2006bw}, we adopt the statistical error of about $25\%$ as uncertainty on the normalization.  This underestimates the total error, that includes a number of systematic effects, some of which are difficult to estimate.  

In principle, to rely on the \sfr\ data is not necessary: the \snr\ is given by direct \sn\ observations.
However, data from \sn\ surveys are sparse (low statistics) and extend only  to $z\sim 1$. The most precise measurement is the recent LOSS result for the local ($z=0$) rate, integrated over all masses~\cite{Li:2010kd}:  $\int_{M_0}^{M_{\rm max} } dM\  \dot{\rho}_{SN}(0,M)  = (0.70 \pm 0.15)\times 10^{-4}~{\rm Mpc^{-3} yr^{-1}}$.  Here we use its error (dominated by systematics), of $\sim 22\%$ as a very optimistic estimate of uncertainty. Other errors are due to possible dust obscuration and supernova impostors (see~\cite{Hopkins:2006bw} for details).

Surprisingly,  the normalization from direct \sn\ observations is lower than that from \sfr\ data, by a factor $\sim 2$ (with a $\sim 2 \sigma$ level~ significance at $z=0$) and by a smaller factor at higher $z$~\cite{Horiuchi:2011zz}.  A possible explanation to this mismatch is that many
supernovae are missed because they are either optically dim (low-luminosity) or dark, whether intrinsically 
or due to obscuration.   Other proposed explanations are an incomplete understanding of 
the star formation and supernova rates including that supernovae form differently in small galaxies than in normal galaxies~\cite{Horiuchi:2011zz}.

Finally, a complementary constraint on the \snr\ normalization comes from the \sk\ limit on the \df\ (Sec.~\ref{sec:dsnbgeneral}). The constraint depends on the \n\ spectrum and is strong for rather hot \n\ spectra~\cite{Strigari:2005hu}. For the cooler spectra favored by SN 1987A, data allow a normalization up to  $\sim 2$ times higher  that the value we use here.

%%%%%%%%%%%%%%%%%%%%%%%%%%%%%%%%%%%%%%%%%%%%%%%%%%%%%%%%%%%%%%%%%%%%%%
\subsection{Reference neutrino signals from supernovae of different masses}                        \label{sec:models}
%%%%%%%%%%%%%%%%%%%%%%%%%%%%%%%%%%%%%%%%%%%%%%%%%%%%%%%%%%%%%%%%%%%%%%
\begin{figure}[t]
\centering
\includegraphics[width=1.\textwidth]{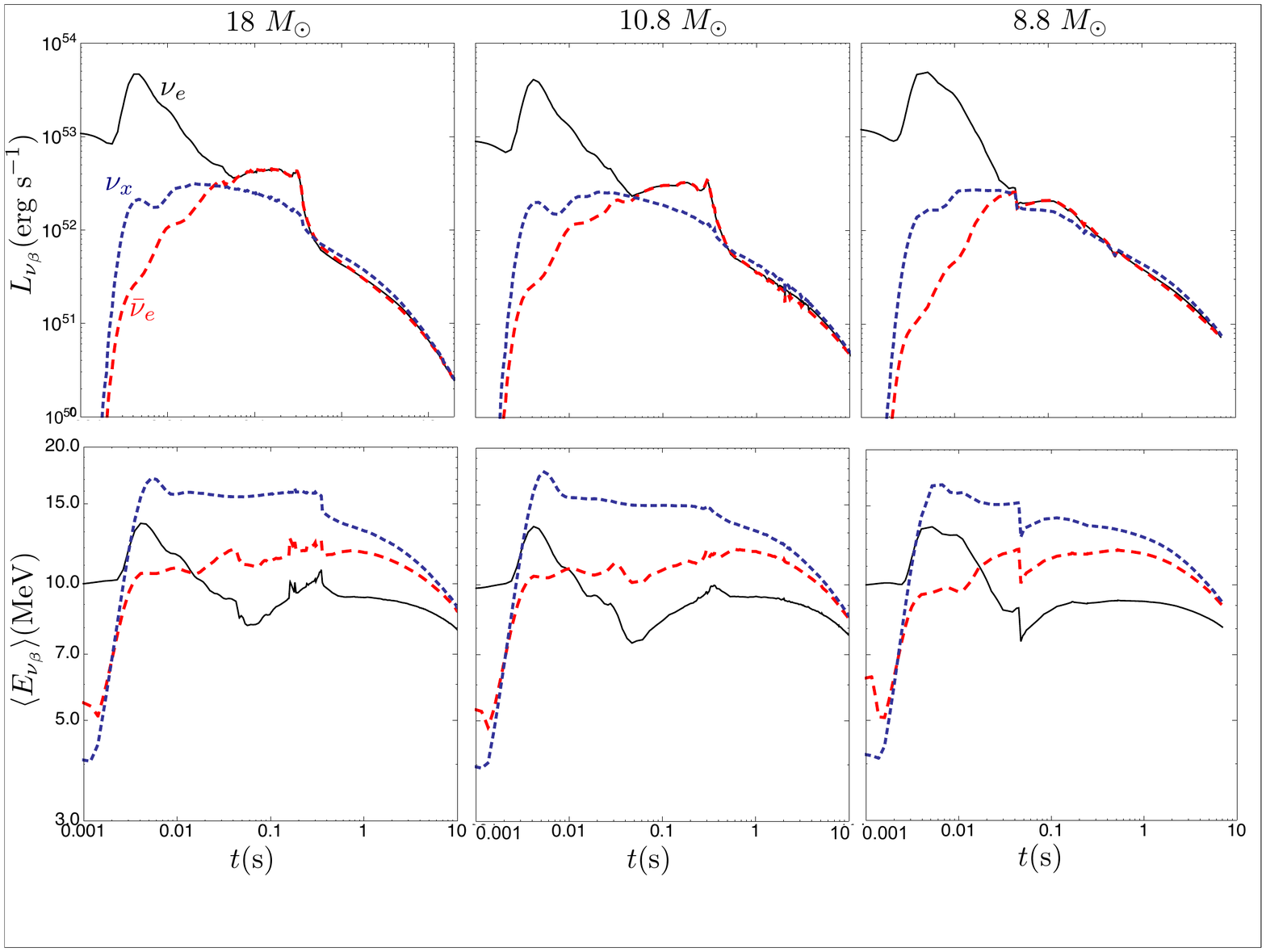}
 \caption{Luminosities (on the top) and mean energies (on the bottom) in the observer frame for three 
 progenitors with masses $M= 18,10.8,8.8\ M_\odot$ (from left to right) as a function of the post-bounce time~\cite{Fischer:2009af,Fischer}. 
 In blue (red, black respectively) are plotted the quantities related to $\nu_{\mu,\tau}$ and $\bar{\nu}_{\mu,\tau}$ denoted by $\nu_x$ ($\bar{\nu}_e$, $\nu_e$). 
}
 \label{lum_meane}
\end{figure}

We adopt supernova  simulations consistently developed over $7$~s or more
for three different masses ($8.8\ M_\odot$, $10.8\ M_\odot$ and $18\ M_\odot$)~\cite{Fischer:2009af}. 
Such core-collapse models  are based on spherically symmetric 
general relativistic hydrodynamics and they include spectral three-flavor Boltzmann neutrino transport with the
equation of state from Shen et al.~\cite{Shen:1998by}. However, they do not include nucleon recoil. 
As observed in~\cite{Huedepohl:2009wh}, the inclusion of nucleon recoil is responsible for reducing the differences among 
the mean energies of different flavors during the cooling phase (otherwise different up to $10\%$)~\cite{Mueller:2012is,Mueller:2011aa,Huedepohl:2009wh}. 
Since one of our purposes is to evaluate the impact of neutrino oscillations on the DSNB, the Basel model favors the appearance  of any 
signature due to neutrino oscillations in the DNSB .

At a radius $r$, the unoscillated spectral number fluxes for flavor
$\nu_\beta$ and for each post-bounce time $t_{\rm pb}$ are 
%.............................................................
\begin{equation}
f_{\nu_\beta}^0(E,t_{\rm pb})= \frac{L_{\nu_\beta}(t_{\rm pb})}{4 \pi r^2}\,\frac{\varphi_{\nu_\beta}(E,t_{\rm pb})}{\langle E_{\nu_\beta}(t_{\rm pb}) \rangle} = \frac{F_{\nu_\beta}^0(E,t_{\rm pb})}{4 \pi r^2} \ ,
\end{equation}
%.............................................................
where $L_{\nu_\beta}(t_{\rm pb})$ is the luminosity for flavor $\nu_\beta$, $\langle
E_{\nu_\beta}(t_{\rm pb}) \rangle$ the mean energy, and $\varphi_{\nu_\beta}(E,t_{\rm pb})$ a quasi-thermal
spectrum. We describe it schematically in the form~\cite{Keil:2002in}
\begin{equation}
\label{alphafit}
\varphi_{\nu_\beta}(E,t_{\rm pb})=\xi_\beta(t_{\rm pb}) \left(\frac{E}{\langle E_{\nu_\beta}(t_{\rm pb}) \rangle}\right)^{\alpha_\beta(t_{\rm pb})} e^{-(\alpha_\beta(t_{\rm pb})+1) E/\langle E_{\nu_\beta}(t_{\rm pb}) \rangle}\ .
\end{equation}
The parameter $\alpha_\beta(t_{\rm pb})$ is defined by $\langle E_{\nu_\beta}(t_{\rm pb})^2
\rangle/\langle E_{\nu_\beta}(t_{\rm pb}) \rangle^2 =
(2+\alpha_\beta(t_{\rm pb}))/(1+\alpha_\beta(t_{\rm pb}))$ and $\xi_\beta(t_{\rm pb})$ is a normalization
factor such that $\int~dE \, \varphi_{\nu_\beta}(E,t_{\rm pb})=1$. 

Figure~\ref{lum_meane} shows the luminosities (on the top) and the mean energies (on the bottom) in the observer frame for the three adopted progenitor
models as a function of the post-bounce time as in~\cite{Fischer:2009af, Fischer}. Note that the luminosities of the different flavors are 
almost equal during the accretion phase, while $L_{\nu_e} \gg L_{\bar{\nu}_e},L_{\nu_{\mu,\tau}}$ during the cooling phase. While, the mean energies are $\langle E_{\nu_{\mu,\tau}}\rangle > \langle E_{\bar{\nu}_e}\rangle > \langle E_{\nu_e}\rangle$ during the cooling phase.

%%%%%%%%%%%%%%%%%%%%%%%%%%%%%%%%%%%%%%%%%%%%%%%%%%%%%%%%%%%%%%%%%%%%%%
\subsection{Neutrino mixing parameters and quantum kinetic equations}		\label{sec:qke}
%%%%%%%%%%%%%%%%%%%%%%%%%%%%%%%%%%%%%%%%%%%%%%%%%%%%%%%%%%%%%%%%%%%%%%

We assume the following neutrino mass squared differences~\cite{Fogli:2011qn}
%...........................
\begin{eqnarray}
\label{masses}
\delta m^2_{\rm atm} &=& 2.35\times 10^{-3}\mathrm{\ eV}^2\ ,\\
\delta m^2_{\rm sol} &=& 7.58\times 10^{-5}\mathrm{\ eV}^2\ ,
\end{eqnarray}
%...........................
and we discuss both normal (NH, $\delta m_{\rm atm}^2>0$) and inverted hierarchy (IH, $\delta m_{\rm atm}^2<0$) scenarios. 
The mixing angles are~\cite{Fogli:2011qn,Fogli:2012ua}
%...................................................
\begin{eqnarray}
\label{theta1312}
\sin^2\theta_{13}=0.02\ \mathrm{and}\ \sin^2\theta_{12}=0.3\ ;
\end{eqnarray}
%....................................................
we neglect the third mixing angle $\theta_{23}$ for reasons that will be clear in a while.

 %........................................

We treat neutrino oscillations in terms of the matrices of
neutrino densities  for each neutrino mode with energy $E$, $\rho_E$, 
where diagonal elements are neutrino densities, off-diagonal
elements encode phase information due to flavor oscillations.
 The radial flavor variation of the
quasi-stationary neutrino flux is given by the ``Schr\"odinger
equation''
%....................................................
\begin{equation}\label{eq:eom1}
\mathit{i}\partial_r\rho_E=[{\sf H}_{E},\rho_{E}]
\quad\hbox{and}\quad
\mathit{i}\partial_r\bar\rho_E=[\bar{\sf H}_{E},\bar\rho_{E}]\,,
\end{equation}
%....................................................
where an overbar refers to antineutrinos and sans-serif letters
denote $3{\times}3$ matrices in flavor space consisting of $\nu_e$,
$\nu_\mu$ and $\nu_\tau$. The initial conditions are
$\rho_{E}=\mathrm{diag}(n_{\nu_e},n_{\nu_\mu},n_{\nu_\tau})$ and $\bar\rho_{E} =
\mathrm{diag}(n_{\bar{\nu}_e},n_{\bar{\nu}_\mu},n_{\bar{\nu}_\tau})$ where $n_{\nu_\mu}=n_{\nu_\tau}=n_{\nu_x}$ and
the same for antineutrinos. The Hamiltonian matrix
contains vacuum, matter, and neutrino--neutrino terms
%....................................................
\begin{equation}
{\sf H}_{E}= {\sf H}^{\rm vac}_{E}+{\sf H}^{\rm m}_{E}+{\sf H}^{\nu\nu}_{E}\ .
\label{eq:ham}
\end{equation}
%....................................................
In the flavor basis, the vacuum term is a function of the mixing
angles and the mass-squared differences
%....................................................
\begin{equation}
{\sf H}^{\rm vac}_{E} = {\sf U}\,\mathrm{diag}\left(-\frac{\omega_{\rm L}}{2},+\frac{\omega_{\rm L}}{2},\omega_{\rm H}\right) {\sf U}^{\dagger}\ ,
\end{equation}
%....................................................
where ${\sf U}$ is the unitary mixing matrix, function of the mixing angles, transforming between
the mass and the interaction basis, $\omega_L = \delta m^2_{\rm sol}/2E$ and $\omega_H = \delta m^2_{\rm atm}/2E$. 
The matter term includes charged-current (CC) interactions and it is responsible for MSW resonances. 
 In the flavor basis, it is 
%....................................................
\begin{eqnarray}
\label{lambda}
{\sf H}^{\rm m} &=& \sqrt{2}G_{\rm F}\;
{\rm diag}(N_{e},0,0)\;,
\end{eqnarray}
%....................................................
where  $N_{e}$ is the net electron number density (electrons minus
positrons).

The corresponding 3$\times$3 matrix caused by neutrino-neutrino
interactions~\cite{Sigl:1992fn} including the multi-angle dependence is
%....................................................
\begin{equation}
{\sf H}^{\nu\nu}_{E} = \int dE' (\rho_{E'}-\bar{\rho}_{E'}) (1-\cos \theta_{EE'})\ ,
\end{equation}
%....................................................
where $\theta_{EE'}$ is the angle between the momenta of the colliding neutrinos~\cite{Duan:2006an}. 

Ignoring radiative corrections, $\nu_\mu$ and $\nu_\tau$
are exactly equivalent in supernovae, allowing us to define an interaction basis 
$(\nu_e, \nu_x,\nu_y) = R^T(\theta_{23})\ (\nu_e,\nu_\mu,\nu_\tau)$, with $R^T(\theta_{23})$ the 
transpose of the rotation matrix $R(\theta_{23})$, such that 
effectively $\theta_{23}=0$~\cite{Dasgupta:2007ws}.
Note that, in this rotated basis, the $e-y$ oscillations are driven by $\delta m^2_{\rm atm}$ while
 the $e-x$ oscillations are driven by $\delta m^2_{\rm sol}$.

It is possible to factorize the effects of self-induced flavor conversions
and the MSW resonances in most of the cases  since they are occurring in well separated regions. We consider multi-angle 
$\nu$--$\nu$ interactions driven by the atmospheric mass difference $\delta m_{\rm atm}$ and by the mixing angle $\theta_{13}$ between
$\nu_e$ and $\nu_y$, while the other flavor ($\nu_x$) does not evolve. In fact the effects on collective flavor conversions induced by 
the third flavor are mainly of the nature of a subtle correction~\cite{Friedland:2010sc,Duan:2010bg,Dasgupta:2010cd, Dasgupta:2010ae} negligible for our purposes.  
The only way $\nu_x$ could affect the final neutrino spectra is by MSW transitions, occurring at larger radii than the 
ones where collective effects happen.
 Thus the fluxes after the collective oscillations  ($F^{c}_{\nu_{\beta}}$) are given by
\begin{eqnarray}
\label{Fce}
F^{c}_{\nu_e}= P_{c}F^{0}_{\nu_e} + (1-P_{c})F^{0}_{\nu_y}\ {\rm and}\ F^{c}_{\bar{\nu}_e}= \bar P_{c}F^{0}_{\bar{\nu}_e} + (1-\bar P_{c})F^{0}_{\bar\nu_y}\ ~,
\end{eqnarray}
 where  $P_c$ and $\bar P_c$ are the $\nue$ and $\barnue$ survival probabilities after the multi-angle self-induced flavor conversions. They are functions of the \n\ energy, and depend on the mass hierarchy. 
The oscillated $\nu_y$ and $\bar \nu_y$ fluxes,  $F^{c}_{\nu_{y}}$ and $F^{c}_{\bar\nu _y}$, can be estimated from the conservation of the total number of $\nu$ (and $\bar{\nu}$):  
\begin{eqnarray}
\label{Fcy}
F^{c}_{\nu _y} + F^{c}_{\nu_e} = F^{0}_{\nu_e} + F^{0}_{\nu_y}\ {\rm and}\  F^{c}_{\bar\nu _y} + F^{c}_{\bar{\nu}_e} = F^{0}_{\bar{\nu}_e} + F^{0}_{\bar{\nu}_y}\ .
\end{eqnarray} 
The third state, $\nu_{x}$ ($\bar\nu_{x}$), is not affected by the collective evolution, therefore  $F^{c}_{\nu _x (\bar\nu _x)}= F^{0}_{\nu_x (\bar\nu_x)}$.

As we consider  the self-induced neutrino oscillations as factorized from the MSW  in first approximation, the  fluxes $F^{c}_{\nu_{\beta}}$ will undergo the traditional MSW conversions after $\nu$--$\nu$ interactions.  In NH, the MSW resonance due to $\delta m^2_{\rm atm}$ affects the $\nu_e$ flux while the $\bar{\nu}_e$ flux remains almost unaffected. 
On the other hand for IH, the same resonance affects the $\bar{\nu}_e$ flux and not the $\nu_e$ flux. 
The fluxes ($F_{\nu_{e}}$ and $F_{\bar \nu_{e}}$) reaching the earth after both the collective and MSW oscillations for NH and IH 
and for large $\theta_{13}$  are~\cite{Chakraboty:2010sz,Dighe:1999bi,Dasgupta:2007ws}:
\begin{eqnarray}
\label{fluxtable1}
F_{\nu_e}^{\rm NH}&=& \sin^2 \theta_{12} [1- P_{c}(F^{c}_{\nu_{e}},F^{c}_{\bar\nu_{e}},E)] (F^{0}_{\nu_e}-F^{0}_{\nu_y})  + F^{0}_{\nu_y}\ , \\ 
F_{\bar{\nu}_e}^{\rm NH}&=& \cos^2 \theta_{12} \bar P_{c}(F^{c}_{\nu_{e}},F^{c}_{\bar\nu_{e}},E) (F^{0}_{\bar{\nu}_e}-F^{0}_{\nu_y}) + F^{0}_{{\nu}_y}\ ,\\ \label{fluxtable2}
F_{\nu_e}^{\rm IH}&=& \sin^2 \theta_{12} P_{c}(F^{c}_{\nu_{e}},F^{c}_{\bar\nu_{e}},E) (F^{0}_{\nu_e}-F^{0}_{\nu_y}) + F^{0}_{{\nu}_y}\ , \\ \label{fluxtable3}
F_{\bar{\nu}_e}^{\rm IH} &=& \cos^2 \theta_{12}  [1-\bar P_{c}(F^{c}_{\nu_{e}},F^{c}_{\bar\nu_{e}},E)] (F^{0}_{\bar{\nu}_e}-F^{0}_{\nu_y}) + F^{0}_{\nu_y}\ .
\label{fluxtable4}
\end{eqnarray}

Here we have used the fact that, by combining Eqs.~(\ref{Fce}, \ref{Fcy}) with $\int dE (F^{0}_{\nu_e} - F^{0}_{\bar \nu_e}) = \mathrm{const.}$, one can express $P_c$ and $\bar P_c$,  for each energy, as functions of the fluxes after collective oscillations:  $P_{c}(F^{c}_{\nu_{e}},F^{c}_{\bar\nu_{e}},E)$ and $\bar P_{c}(F^{c}_{\nu_{e}},F^{c}_{\bar\nu_{e}},E)$.   These probabilities exhibit a well known step-like behavior, that appears in the fluxes  $F_{\nu_{e}}$ and $F_{\bar \nu_{e}}$ as the so called ``spectral splits"~\cite{Fogli:2007bk,Fogli:2009rd}. In reality, a number of effects smooth out the splits in the observed \n\ fluxes; we discuss this point further below.

%%%%%%%%%%%%%%%%%%%%%%%%%%%%%%%%%%%%%%%%%%%%%%%%%%%%%%%%%
\section{Time-integrated neutrino fluxes and oscillation effects}		\label{sec:models}
%%%%%%%%%%%%%%%%%%%%%%%%%%%%%%%%%%%%%%%%%%%%%%%%%%%%%%%%%%%%%%%%%%%%%%
\begin{figure}[t]
\centering
\includegraphics[width=1\textwidth]{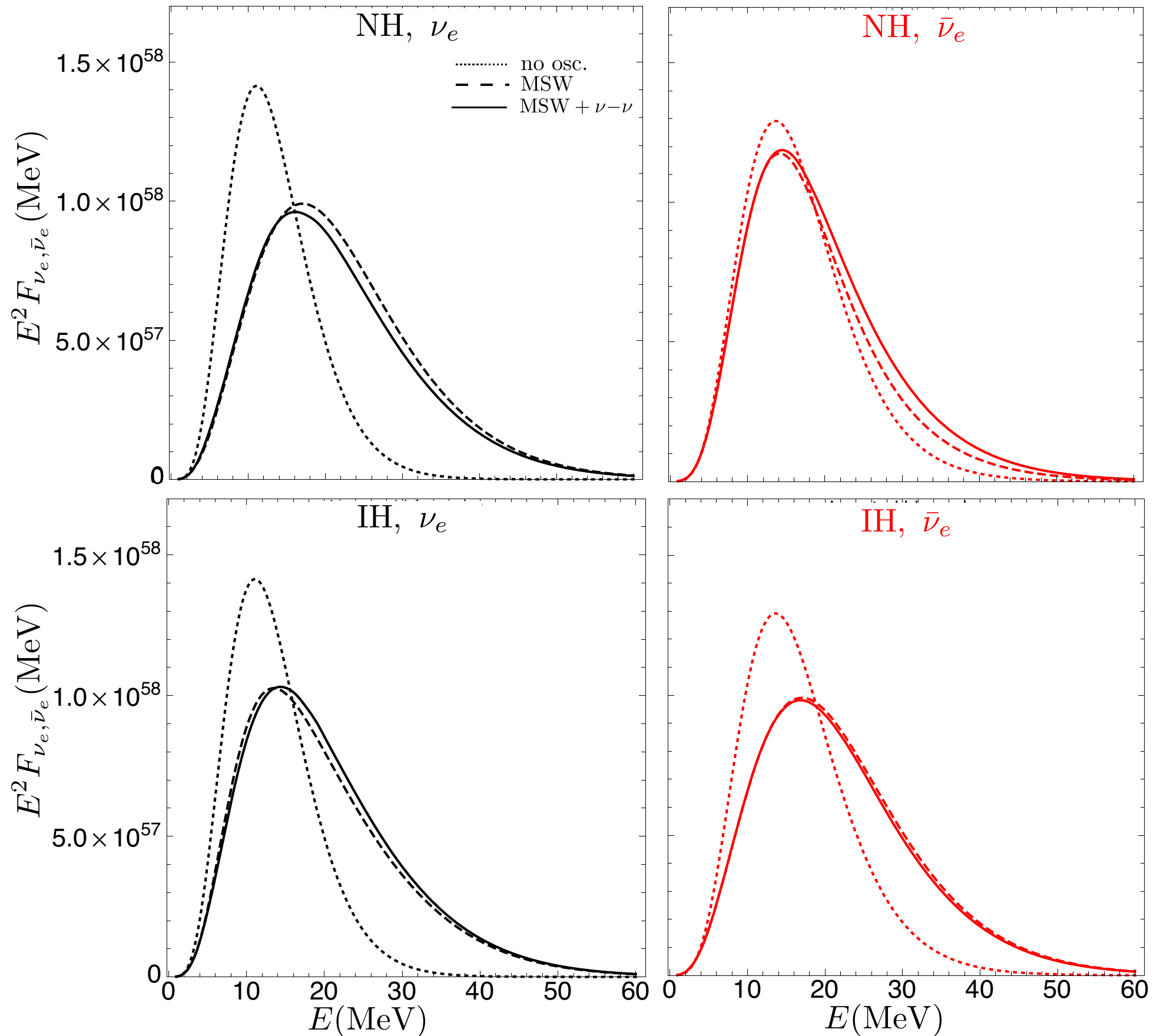}
 \caption{Time-integrated $E^2 \times F_{\nu_\beta}(E)$ fluxes as a function of the energy for $M=10.8\ M_\odot$, for $\nu_e$ ($\bar{\nu}_e$) 
 on the left (right) and NH (IH) on the top(bottom). The non-oscillated fluxes are plotted  with dotted lines, the fluxes obtained including neutrino interactions
 with matter only (MSW) are plotted with dashed lines and the fluxes obtained including both matter and $\nu$ background (MSW+$\nu$--$\nu$) are plotted
 with solid lines.}
 \label{timeint_flux}
\end{figure}

As a first step towards the calculation of the \df, we compute the oscillated fluxes, Eqs. (\ref{fluxtable1})-(\ref{fluxtable4}), for each progenitor model, at fixed time snapshots (as an approximation of the continuous time evolution, which is too demanding for state of the art computers). The results are then used to obtain the $\nu_\beta$ and $\bar \nu_\beta$  fluxes integrated over the duration of the \n\ burst.  For illustration, in this section we discuss the results for the $10.8 \msun$ progenitor; in Sec.~\ref{sec:dsnb} the fluxes for all progenitors will be summed up to obtain the \df, via Eq.~(\ref{phiDSNB}).  

While the MSW effect is well described analytically, the collective effects require a numerical calculation to obtain the probabilities $P_c$ and $\bar P_c$.  Let us discuss them here in more detail. 

The spectral split patterns (affecting $P_c$ and $\bar P_c$) are known to be crucially dependent on the initial relative flux densities and on the mass hierarchy. For definiteness, it is convenient to distinguish between the probabilities in the accretion phase ($t_{\rm pb} \le 1$~s), and those in the cooling phase ($t_{\rm pb} \gta 1$~s).  
In  the accretion phase, the multi-angle effects associated with dense ordinary matter suppress  collective effects~\cite{Chakraborty:2011nf,Chakraborty:2011gd,Saviano:2012yh,Sarikas:2011am}. 
Therefore, we adopt the results presented in~\cite{Chakraborty:2011gd,Saviano:2012yh} for the two-flavor system $(\nu_e,\nu_y)$, considering partial or no flavor conversion for several $t_{\rm pb}$ as in~\cite{Chakraborty:2011gd,Saviano:2012yh}. 

During the cooling phase, the fluxes of different flavors are slightly different,  and 
spectral splits occur for neutrinos and/or antineutrinos according to the mass hierarchy 
and to the number of crossings in the non-oscillated spectra (i.e., energies where $F_{\nu_e}^0(E)=F_{\nu_x}^0(E)$ and 
the same for $\bar{\nu}$)~\cite{Dasgupta:2009mg}.  We calculate these effects by numerically solving Eqs.~(\ref{eq:eom1}) for $t_{\rm pb}=1,3,6,9$~s, for the system $(\nu_e,\nu_y)$.   Concerning the neutrino emission geometry, we assume a spherically symmetric source emitting neutrinos and 
antineutrinos like a blackbody surface, from a neutrinosphere with radius, $R_\nu$, that varies   with $t_{\rm pb}$. We define the 
neutrinosphere radius as the radius where the neutrino radiation field is half-isotropic~\cite{EstebanPretel:2007ec,Chakraborty:2011gd}
and we adopt the same $R_\nu$ for different flavors. 
Specifically, we  use  $\delta m^2_{\rm atm}$ as in Eq.~(\ref{masses}), $R_{\nu} \simeq 13$~km (although slightly varying with $t_{\rm pb}$) and, to be coherent with the treatment in~\cite{Chakraborty:2011gd,Saviano:2012yh}, we simulate the matter effect on collective oscillations using a small, effective, in-medium mixing angle $\theta_{13, {\rm{eff}}} = 10^{-3}$~\cite{EstebanPretel:2008ni}. The  non-oscillated spectra are defined as in Eq.~(\ref{alphafit}) with mean energies and 
luminosities as in Fig.~\ref{lum_meane}.

We find that that multiple crossings  appear in the non-oscillated spectra and therefore 
multiple spectral splits are expected during the cooling phase due to collective effects.
However, the shape of the splits is smeared and their size is reduced due to the similarity of the 
unoscillated flavor spectra and to multi-angle effects. Often, only partial conversion is realized.
  In particular, for NH we find that  $P_c$ exhibits a transition from high ($P_c \gta 0.6$) to low ($P_c \lta 0.3$) values as the energy increases, with transition energy increasing with time from $\sim 15$ to $\sim 35$ MeV. A similar behavior is observed for  $\bar P_c$ for energies of relevance for detection, $E\gta 5-7 $ MeV.  For IH, $P_c$ has a somewhat opposite energy dependence,  changing from $P_c \sim 0.1-03$ at $E \lta 15-20$ MeV to $P_c \gta 0.6$ at higher energy.  The transition becomes gentler in energy as time increases.  For antineutrinos, we have $\bar P_c \gta 0.7$ at all times and all energies, with only a moderate energy dependence. 
  Our results for  $P_c, \bar P_c$ are qualitatively similar to what shown in previous literature, although smoothened by the multi-angle dependence and depending on the adopted supernova progenitor; therefore we choose to be brief here, and refer to dedicated papers for details~\cite{Fogli:2009rd,Mirizzi:2010uz}. 

To obtain the time-integrated \n\ fluxes, we approximate the time dependence of $P_c$ and $ \bar P_c$ with a step-like form,  taking them to be constant in  time intervals centered on the scanned $t_{\rm pb}$ (we adopt a time window of $\sim 3$~s centered on each $t_{\rm pb}$). We checked that the particular choice of the time-binning affects the final results only weakly.  The time-integrated fluxes have been obtained by integrating over time the fluxes computed at each time slice.

Figure~\ref{timeint_flux} shows the time-integrated fluxes, $E^2 \times F_{\nu_\beta}(E)$, as a function of the energy for $\nu_e$ ($\bar{\nu}_e$) on the left (right),  for both mass hierarchies for the $10.8\ M_\odot$ supernova model. We distinguish among
non-oscillated integrated fluxes (dotted lines), oscillated fluxes including MSW effects only (dashed lines) and oscillated fluxes including both
MSW and $\nu$--$\nu$ interactions (solid lines).  
Note that the MSW effect is large and generates hotter fluxes. Compared to MSW only, the collective effects produce a slight hardening or softening of 
the fluxes depending on the hierarchy, according to Eqs.~(\ref{fluxtable1}--\ref{fluxtable4}).   
No signature of multiple spectral splits appear in the time-integrated flux, because they are suppressed during the accretion and they occur at different energies for each $t_{\rm pb}$ during the cooling; moreover multi-angle effects smear the conversion probabilities.  
 
Overall, the flux variation due to collective effects -- relative to MSW conversion only -- is at the level of 10\% or less.  Such a small difference is explained by the fact that, in contrast with earlier approximate results on collective effects, the probabilities $P_c $ and $\bar P_c$ rarely approach zero, but remain in the interval $\sim 0.3-0.8$ in most cases. One should also consider that the contribution of $P_c $ and $\bar P_c$ appears in Eqs.~(\ref{fluxtable1}--\ref{fluxtable4}) weighed by factors of $\sin^2\theta_{12}$ or    $\cos^2\theta_{12}$, that reduce its size.   A further reduction takes place  in the time-integrated flux, which receives about 1/2 of its luminosity from the accretion phase, where collective effects are suppressed.  Finally, 
the differences between the non-oscillated fluxes in the different flavors become progressively smaller in the cooling phase, thus suppressing any spectral change due to oscillations.

To further illustrate the size of each effect we included, in Tab.~\ref{fittable} we provide  the mean energies and the mean squared energies for the time-integrated
spectra. In the existing literature, one progenitor has been often considered 
as representative of the whole stellar population and one post-bounce time was taken as representative of both accretion and cooling phase. 
Therefore, for sake of completeness, we give this case (case a in the Table) using the parameters for $t_{\rm pb}=0.5$~s as representative of all post-bounce 
times and the $10.8\ M_\odot$ progenitor as representative of the whole stellar population. 
Looking at the average energies in the Table, we see an expected  decrease (5-10\% or so) when the full cooling over $\sim 10$ s is included (case b), and a marked hardening due to including the MSW conversion  (up to 30\%, case c).  As expected from Fig.~\ref{timeint_flux}, adding collective effects amounts to a change of 10\% or less (case d), and  only a very minor hardening is due to summing the results for different progenitor masses (case e). In the next Section, we will discuss in detail the differences among the different cases for the \df.

%%%%%%%%%%%%%%%%%%%%%%%%%%%%%%%%%%%%%%%%%
\begin{table}[t]
\center
\begin{tabular}{llllll}
\hline
 &a & b & c & d & e \\
 \hline
$\langle E_{\nu_e} \rangle_{\rm NH}$~(MeV)  &$9.53$ & $8.96$ & $12.24$ & $11.90$& $11.96$\\
$\langle E_{\nu_e}^2 \rangle_{\rm NH}$~(MeV$^2$)  &$113.85$ & $100.18$ & $209.53$ & $198.13$& $199.66$\\
$\langle E_{\bar{\nu}_e} \rangle_{\rm NH}$~(MeV)  &$11.82$ & $10.84$ & $11.23$ & $11.50$& $11.62$\\
$\langle E_{\bar{\nu}_e}^2 \rangle_{\rm NH}$~(MeV$^2$)  &$174.20$ & $149.17$ & $165.88$ & $176.72$ &$179.62$\\
\hline
\hline
$\langle E_{\nu_e} \rangle_{\rm IH}$~(MeV)  &$9.53$ & $8.96$ & $10.83$ & $11.08$& $11.23$\\
$\langle E_{\nu_e}^2 \rangle_{\rm IH}$~(MeV$^2$)  &$113.85$ & $100.18$ & $162.56$ & $170.62$& $174.48$\\
$\langle E_{\bar{\nu}_e} \rangle_{\rm IH}$~(MeV)  &$11.82$ & $10.83$ & $12.24$ & $12.17$ &$12.18$\\
$\langle E_{\bar{\nu}_e}^2 \rangle_{\rm IH}$~(MeV$^2$)  &$174.20$ & $149.17$ & $209.53$ & $206.77$ &$206.06$\\
\hline
\end{tabular}
\caption{Mean energies and mean squared energies for the the time-integrated $F_{\nu_\beta}(E)$. The different cases considered are: (a) $F_{\nu_\beta}(E)$ obtained considering $t_{\rm pb}=0.5$~s as representative of all post-bounce times and the $10.8\ M_\odot$ progenitor as representative of the whole stellar population, no oscillations; (b)  $F_{\nu_\beta}(E)$ obtained considering time-dependent fluxes for the $10.8\ M_\odot$ model, no flavor oscillations; (c) $F_{\nu_\beta}(E)$ obtained considering time-dependent fluxes for the $10.8\ M_\odot$ model, MSW flavor conversions; (d) $F_{\nu_\beta}(E)$ obtained considering time-dependent fluxes for the $10.8\ M_\odot$ model, MSW + $\nu$--$\nu$ interactions; (e) $F_{\nu_\beta}(E)$ obtained including the whole stellar population, time-dependent fluxes, MSW + $\nu$--$\nu$ interactions.}
\label{fittable}
\end{table}
%%%%%%%%%%%%%%%%%%%%%%%

%%%%%%%%%%%%%%%%%%%%%%%%%%%%%%%%%%%%%%%%%%%%%%%%%%%%%%%%%%%%%%%%%%%%%%
\section{Diffuse supernova neutrino background} 					\label{sec:dsnb}
%%%%%%%%%%%%%%%%%%%%%%%%%%%%%%%%%%%%%%%%%%%%%%%%%%%%%%%%%%%%%%%%%%%%%% 
\begin{figure}[t]
\centering
\includegraphics[width=1.\textwidth]{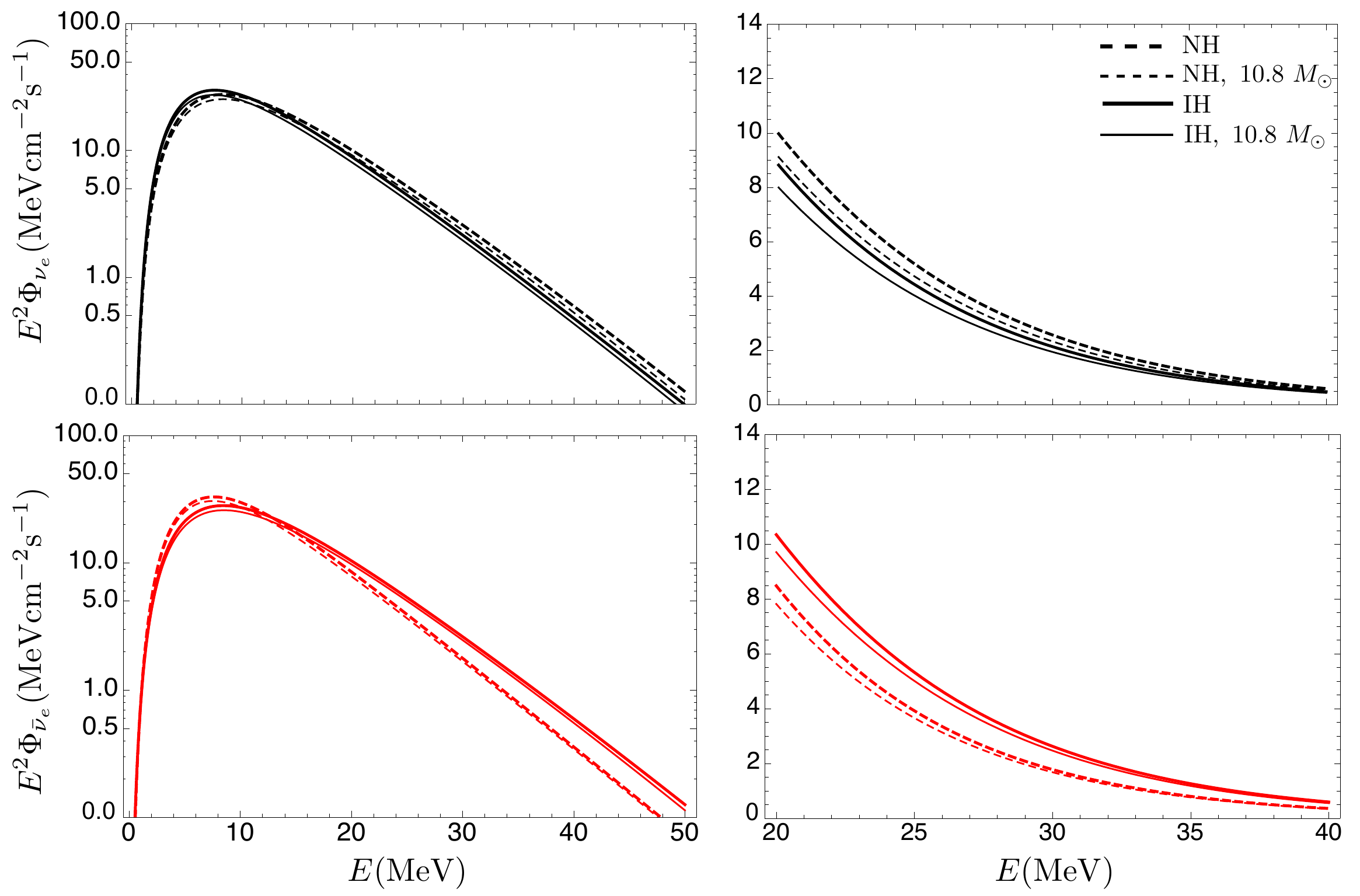}
 \caption{DSNB as a function of the energy ($E^2 \times \Phi_{\nu_e,\bar{\nu}_e}$) on the left panels, zoom in the region of Super-Kamiokande detection in the right panels ($E > 17.3$~MeV). The $\nu_e$ ($\bar{\nu}_e$) flux is plotted in black (red). The solid (dashed) line represents the DSNB in IH (NH). To mark the effects induced by the inclusion of different SN model, the thick lines represent the DSNB including the three different SN progenitors and the thin line is the DSNB obtained extending $10.8\ M_\odot$ to the whole stellar population.}
 \label{DSNB_plot}
\end{figure}

 After computing the time-integrated spectra for each of the considered SN progenitors as explained in Sec.~\ref{sec:models},  we derive the 
DSNB, $\Phi_{\nu_\beta}$,  using Eq.~(\ref{phiDSNB}).  We adopt the $8.8\ M_\odot$ SN as representative of supernovae with mass in the interval $[8,10]\ M_\odot$, the $10.8\ M_\odot$ SN for masses in the interval $[10,15]\ M_\odot$ and the $18\ M_\odot$ progenitor for $M > 15 M_\odot$. Note that the for the first interval we adopt  the widest range expected for O-Ne-Mg core formation.

Figure~\ref{DSNB_plot} shows the DSNB, $E^2 \times \Phi_{\nu_\beta}$ with $\Phi_{\nu_\beta}$, as a function of the energy for $\nu_e$ ($\bar{\nu}_e$) on the top (bottom). 
 In the plots on the right, we focus on the region of interest for Super-Kamiokande 
detection ($E > 17.3$~MeV) and it is possible to appreciate a maximum variation of $10$--$20\%$ (at $E \simeq 20$~MeV) related to the 
mass hierarchy.  
In order to favor a comparison with the existing literature, we show the fluxes obtained by taking the $10.8\ M_\odot$ 
progenitor as representative of all progenitors (thin curves), and the full result that includes the variation of the fluxes 
and probabilities with the progenitor mass (thick lines). Due to the similarity of the initial luminosities and mean energies (see Sec.~\ref{sec:models}) the differences between the two cases are of few per cent.

Shock effects in Fe-core supernovae manifest themselves during the cooling phase and at
large radii. We choose to neglect such effects for the $18\ M_\odot$ and $10.8\ M_\odot$ models because
they practically might affect the $\delta m^2_{\rm atm}$-driven resonance only for the late cooling phase having a negligible impact on the DNSB. On the other hand, the $8.8\ M_\odot$ model has a Ne-O-Mg core 
and the first hundreds of milliseconds of the burst are affected by variations of the matter density profile
due to the shock passage. The shock wave is responsible for turning the flavor conversions
 from non-adiabatic to adiabatic within the first $\sim 250$~ms affecting the $\delta m^2_{\rm atm}$-driven resonance~\cite{Lunardini:2007vn}. However, a very small time-window 
 would have been affected by the   shock passage in the electron-capture SN and we expect its impact on the total DSNB to be negligible.

Recently, it has been observed that the non-forward neutrino scattering contribution might affect flavor conversions 
during the accretion phase~\cite{Cherry:2012zw,Sarikas:2012vb}. These studies reveal that  the consequences of non-forward
scatterings are negligible for large SN masses when complete matter suppression occurs~\cite{Sarikas:2012vb} and for
low mass supernovae as the Ne-O-Mg ones~\cite{Cherry:2012zw}. However, it might be still relevant  for supernovae 
with intermediate masses (as the $10.8\ M_\odot$ model).  
Since these studies are still embryonal and no numerical implementation of the non-forward 
contribution is available yet, a priori, one could expect larger flavor conversions induced by the non-forward term 
during the accretion phase especially for the intermediate SN mass range. Therefore, we chose  
to adopt a parametric study of the accretion phase for the $10.8\ M_\odot$ model, considering two extreme cases of full flavor conversions ($P_c=0$)
and complete matter suppression ($P_c=1$) due to multi-angle matter effects, other than the intermediate scenario 
presented in~\cite{Saviano:2012yh,Chakraborty:2011gd,Chakraborty:2011nf}.  We find that, at the level of the DSNB, 
after summing the fluxes from progenitor of different masses, the difference between the two cases (not shown here) is of the order of few per cent.

\begin{figure}[t]
\centering
\includegraphics[width=1.\textwidth]{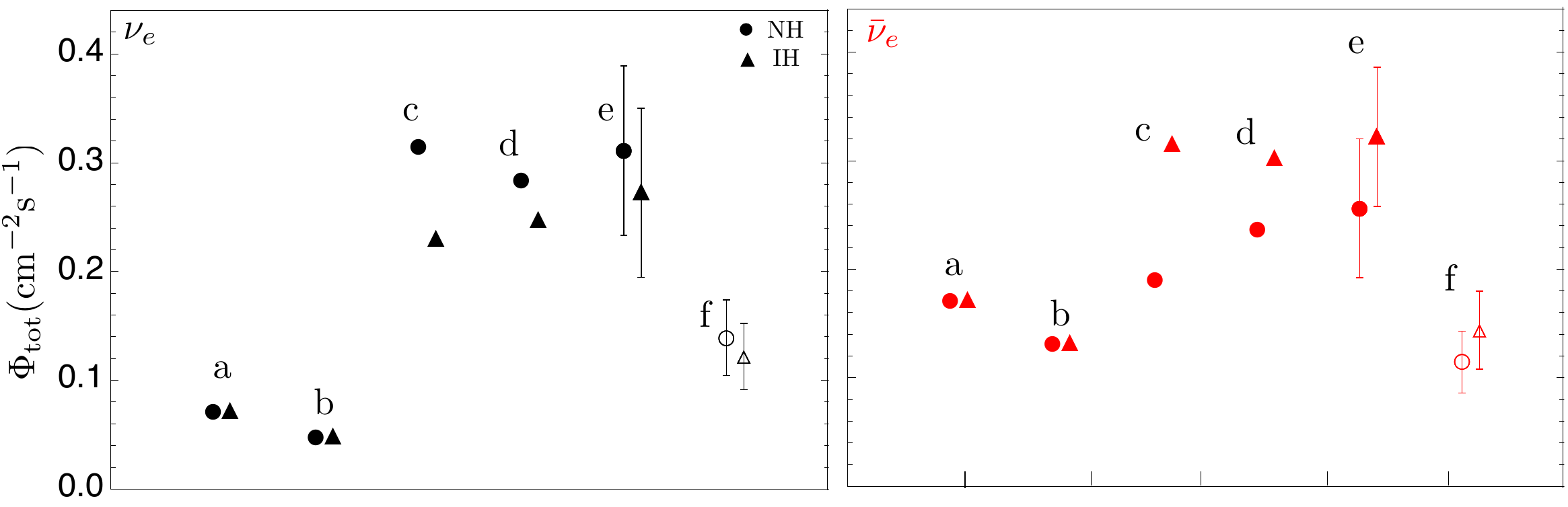}
 \caption{Energy integrated DSNB (for $E > 17.3$~MeV) as a function of the discussed contributions, in black (red) for $\nu_e$ ($\bar{\nu}_e$). 
 We adopt triangles (dots) for the IH (NH) scenario. From left to right the source of variation of the DSNB are: (a) DSNB  for $t_{\rm pb}=0.5$~s as representative of all post-bounce 
times and the $10.8\ M_\odot$ progenitor as representative of the whole stellar population, no oscillations; (b) DSNB obtained considering time-dependent fluxes for the $10.8\ M_\odot$ model, no flavor oscillations; (c) DSNB obtained considering time-dependent fluxes for the $10.8\ M_\odot$ model, MSW flavor conversions; (d) DSNB obtained considering time-dependent fluxes for the $10.8\ M_\odot$ model, MSW + $\nu$--$\nu$ interactions; (e,f) DSNB obtained including the whole stellar population, time-dependent fluxes, MSW + $\nu$--$\nu$ interactions; the difference in the two points is due to the factor of $\sim 2$ between the results of \sn\ surveys and those of star formation measurements (see Sec.~\ref{sec:CCSFR}). The error bars are the errors on the normalization of the supernova rate~\cite{Horiuchi:2011zz}.}
\label{uncert}
\end{figure}

Figure~\ref{uncert}  gives results for the \df\ integrated in energy above the experimental threshold 17.3 MeV~\cite{Bays:2011si}. It shows results for the cases (a-e) of Tab.~\ref{fittable} for both mass hierarchies. 
The differences between these cases reflect the results found for an individual \sn\ burst in Sec.~\ref{sec:models}, with more emphasis on the high energy part of the \n\ spectrum (due to our high integration threshold). 
Compared to static spectra (case a), the effect of the time-dependence of the spectra over the  $10$~s of the
neutrino burst is responsible for a variation of  $6 \%$ (case b). The MSW effects (case c) are the largest source of variation 
of the DSNB with respect to the case without oscillations and it is about $50$--$60 \%$. This variation is particularly evident for  neutrinos and the NH, where the difference between the unoscillated and oscillated spectra is the largest, due to the complete flavor permutation driven by the large $\theta_{13}$. Neutrino-neutrino interactions (case d) are responsible for a variation of  $5$--$10 \%$ with
respect to the MSW only case. Summing over the stellar population (case e) is responsible for
a DSNB variation of  $5$--$10 \%$ due to the more luminous 
fluxes of the more massive stars. 
For sake of completeness, we provide the numerical values for the DSNB for the case (e):
\begin{eqnarray}
\Phi_{\rm tot}^{\nu_e, \rm NH} = 0.31~{\rm cm}^{-2}{\rm s}^{-1}\ {\rm and}\ \Phi_{\rm tot}^{\nu_e, \rm IH} = 0.27~{\rm cm}^{-2}{\rm s}^{-1}\ , \\ \nonumber
 \Phi_{\rm tot}^{\bar{\nu}_e, \rm NH} = 0.26~{\rm cm}^{-2}{\rm s}^{-1}\ {\rm and}\ \Phi_{\rm tot}^{\bar{\nu}_e, \rm IH} = 0.32~{\rm cm}^{-2}{\rm s}^{-1}\ .
\end{eqnarray}
The maximum impact given by the mass hierarchy is  $20 \%$, triangle and dot in (e). It is realized for antineutrinos because, for large 
$\theta_{13}$, the high density MSW resonance is adiabatic and the $\bar{\nu}_e$ flux changes from almost complete survival for NH 
(survival probability $\sim 0.7$--$0.8$) to almost complete conversion for  IH (survival probability  $\sim 0.1$--$0.2$).
 We estimate the DSNB taking into account the unexplained factor of $\sim 2$ between the results of \sn\ surveys and those of star formation measurements as discussed in Sec.~\ref{sec:CCSFR}, points (e) and (f). This mismatch is responsible for the largest astrophysical source of error on the estimation of the DSNB, $\sim 50\%$. Moreover, the error on the normalization of the supernova rate is about $25 \%$ and it is represented by the error bars in (e) and (f). However, this error could be most likely higher once several systematic 
errors are included.

Another potential source of error on the DSNB is in the  equation of state of nuclear matter used in core collapse simulation (assumed fixed in all our computations). 
The variation of the total neutrino energy release during the supernova explosion reflects the variation of
the gravitational binding energy of the neutron star in dependence of different nuclear equations
 of state (see~\cite{Lattimer:2000nx} for details). However, we expect that the differences of the time-integrated 
 emitted neutrino spectra using different equations of state will be not large enough to play a role in comparison
 will the other astrophysical uncertainties discussed in this paper.

%%%%%%%%%%%%%%%%%%%%%%%%%%%%%%%%%%%%%%%%%%%%%%%%%%%%%%%%%%%%%%%%%%%%%%
\section{Conclusions}
\label{sec:conclusions}
%%%%%%%%%%%%%%%%%%%%%%%%%%%%%%%%%%%%%%%%%%%%%%%%%%%%%%%%%%%%%%%%%%%%%%

The diffuse supernova neutrino background is an integrated picture of all the core collapse \sne\ of different progenitor mass and redshift, and has realistic chances to be detected in the near future by upcoming experiments.  The predictions for this flux have been constantly updated to include new understanding of \sn\ physics and astrophysics, and of \n\ phenomenology.  We present one such update, to include, as consistently as possible, the results of new supernova simulations and oscillation calculations.

Specifically, our calculation is the first that uses \n\ fluxes that are numerically calculated over the entire $\sim 10$ s of the \n\ emission for a set of different progenitor masses.  For each progenitor star, we study neutrino oscillations at different times post bounce, for both the neutrino mass hierarchies, including multi-angle collective effects and adopting the recently measured value of $\theta_{13}$.  The results are then used to obtain the time-integrated flux from a single \sn; finally, the oscillated fluxes for different progenitor masses are combined together into a prediction for the \df.  

The results of this paper rely on the Basel simulation of \n\ emission~\cite{Fischer:2009af}.  
Although model dependent,  they are indicative of what might be expected from state of the art simulations.  In particular, the Basel model has the largest spectral differences between the \n\ flavors, and therefore it is useful to estimate the largest oscillation effects that can be realistically expected.

Let us summarize our main results.
\begin{itemize}
\item The inclusion of time-dependent \n\ spectra are responsible for a colder \n\ spectrum compared to calculations using a fixed spectrum from early time emission (e.g., $t_{\rm pb} \sim 0.5$ s).  For a single supernova, the effect is a $\sim 5\%$ shift in the average energies (Tab.~\ref{fittable}), which translates into a $\sim 5\%$ difference in the integrated \df\ above a 17.3 MeV threshold (Fig.~\ref{uncert}). 
\item The largest effect  of flavor oscillations  is due to the matter-driven MSW resonances ($\sim 50$--$60 \%$), while \n-\n\ collective effects give a further $\sim 5$--$10 \%$ contribution. Moreover, the  DSNB does not present any energy-dependent signature of the collective oscillations. 
For fixed SN progenitor, 
multi-angle matter effects suppress collective oscillations in the accretion phase, while multiple spectral splits occur in the cooling phase. However, such  
splits occur at different energies for different post-bounce times and are smeared by multi-angle effects, disappearing in the DSNB   by the integration over the time and over the \sn\ population. 
\item The dependence on the  mass hierarchy is of the order $10-20\%$  (Figs.~\ref{DSNB_plot} and \ref{uncert}) and it is strongest for antineutrinos.  This is because, with the recently measured large value of $\theta_{13}$, the high density MSW resonance is adiabatic and $\bar{\nu}_e$ change from almost complete survival (survival probability  $\sim 0.7-0.8$) to almost complete conversion (survival probability  $\sim 0.1-0.2$) as the mass hierarchy changes from normal to inverted.  
\item  Combining results for different progenitor stars, instead of using the $10.8 ~\msun$ spectra for all stars, tends to increase the \df\ by  $\sim 5-10\%$, due to the more luminous fluxes of the more massive stars.   
\end{itemize}

In summary, we find that, in first approximation, the \df\ can be described by a simplified model with a fixed, time independent spectrum for all \sne, with MSW oscillations. The error due to neglecting the other effects that we have studied here is of the order of ten per cent or less and, assuming that
the sign of the mass hierarchy will be known within the next decade and the picture of neutrino oscillations in supernovae will not drastically change in the next future, such error will be only related to the modeling of neutrino emission.   We note that this error is smaller than astrophysical uncertainties: the error on the normalization of the supernova rate alone is at least $25 \%$, and most likely higher once several systematic errors are included (among these, the still mysterious factor of $\sim 2$ between the results of \sn\ surveys and those of star formation measurements).  In addition to these, one should also consider errors at \n\ detectors, especially those due to the high level of background.  

In consideration of errors and experimental challenges, how likely is it that the effects we have discussed will have any relevance in future data analyses?  When the \df\ will be detected, probably a first phase of data analysis will focus on excluding a number of models of  \n\ spectra and \snr.  In this respect, to model the \df\ as accurately as possible will be important to correctly establish or exclude compatibility.  To actually test for effects at the 10\% level will require long term progress to reduce at least some of the uncertainties of experimental and theoretical nature. For example, with new, precise measurements of the \snr\ and of  the sign of the  \n\ mass hierarchy from beams, and reliable \n\ spectra from future simulations (or from a galactic \sn),  it might be possible to use the \df\ to tests the luminosity of the \n\ fluxes from lesser known \sn\ with high mass progenitors.

%%%%%%%%%%%%%%%%%%%%%%%%%%%%%%%%%%%%%%%%%%%%%%%%%%%%%%%%%%%%%%%%%%%%%%
\section*{Acknowledgments} %%%%%%%%%%%%%%%%%%%%%%%%%%%%%%%%%%%%%%%%%%%
%%%%%%%%%%%%%%%%%%%%%%%%%%%%%%%%%%%%%%%%%%%%%%%%%%%%%%%%%%%%%%%%%%%%%%
We are grateful to Tobias Fisher for providing the supernova data and for useful discussions. 
We also thank Georg Raffelt for valuable discussions concerning this work and comments on the manuscript.
This work was partly supported by the Deutsche Forschungsgemeinschaft under grant EXC-153, by the  European Union FP7 ITN
INVISIBLES (Marie Curie Actions, PITN-GA-2011-289442) and by the NSF grant PHY-0854827.
 I.T. acknowledges support from the Alexander von Humboldt Foundation and thanks Arizona State University for 
 hospitality during the final stages of this work.
 %%%%%%%%%%%%%%%%%%%%%%%%%%%%%

%%%%%%%%%%%%%%%%%%%%%%%%%%%%%%%%%%%%%%%%%
%\section*{References}
%%%%%%%%%%%%%%%%%%%%%%%%%%%%%%%%%%%%%%%%%%%%%%%%%%%%%%%%%%%%%%%%%%%%%%

\end{document}